\documentclass[12pt]{iopart}

\usepackage{graphicx}
\usepackage{titlesec}
\usepackage{xcolor}
\definecolor{red2}{RGB}{209,35,42}
\definecolor{blue2}{RGB}{11,125,180}
\titlelabel{\thetitle.  \quad \quad \quad}
\titleformat{\section}   %section properties
{\color{red2}\normalfont\large\bfseries} % color , font style, size, bold
{\color{red2}\thesection.}{0.25em}{} %The section number format, color and space between the section number and section name

\begin{document}

\title[Magnetism in Two-Dimensional Ilmenenes]{Magnetism in Two-Dimensional Ilmenenes: \\ Intrinsic Order and Strong Anisotropy}

\author{R.H Aguilera-del-Toro$^{1-2\ast}$, M. Arruabarrena$^{2}$, A. Leonardo$^{1-3}$, A. Ayuela$^{1-2}$}

\address{$^{1}$Donostia International Physics Center (DIPC), 20018 Donostia, Spain}
\address{$^{2}$Centro de F\'isica de Materiales - Materials Physics Center (CFM-MPC), 20018 Donostia, Spain}
\address{$^{3}$EHU Quantum Center, Universidad del País Vasco/Euskal Herriko Unibertsitatea UPV/EHU, Leioa, Spain}
\ead{rodrigo.aguilera@dipc.org
}
\vspace{10pt}
\begin{indented}
\item[]October 2022
\end{indented}

\begin{abstract}
Iron ilmenene is a new two-dimensional material that has
recently been exfoliated from the naturally-occurring iron titanate found in ilmenite ore, a material that is abundant on
earth surface. In this work, we theoretically investigate the
structural, electronic and magnetic properties of 2D
transition-metal-based ilmenene-like titanates. The study of
magnetic order reveals that these ilmenenes usually
present intrinsic antiferromagnetic coupling between the 3d
magnetic metals decorating both sides of the Ti-O layer.
Furthermore, the ilmenenes based on late 3d brass metals, such
as CuTiO$_3$ and ZnTiO$_3$, become ferromagnetic and spin
compensated, respectively. Our calculations
including spin-orbit coupling reveal that the magnetic ilmenenes
have large magnetocrystalline anisotropy energies when the 3d
shell departs from being either filled or half-filled, with
their spin orientation being out-of-plane for elements below
half-filling of 3d states and in-plane above. These
interesting magnetic properties of ilmenenes make them useful
for future spintronic applications because they could be
synthesized as already realized in the iron case.
\end{abstract}

%%%%%%%%%%%%%%%%%%%%%%%%%%%%%%%%%%%%%%%%%%%%% 
%%%%%%%%%%%%%%%% INTRODUCTION %%%%%%%%%%%%%%% 
%%%%%%%%%%%%%%%%%%%%%%%%%%%%%%%%%%%%%%%%%%%%%

\section{\color{red2}Introduction}

Technology for synthesizing two-dimensional materials has greatly improved in recent years. Since the synthesis of graphene\cite{grafeno1,grafeno2}, a large number of extensive systems only a few atoms thick have been obtained. 
The study of these 2D materials has brought new physical phenomena with countless applications into play, like their magnetic properties\cite{gong2017discovery,huang2017layer}. 
Obtaining magnetism in 2D isotropic crystals is forbidden in the Heisenberg model as explained by the Mermin-Wagner theorem\cite{Mermin}: 
the magnon dispersion is reduced with respect to their 3D counterparts, and it has an abrupt onset, which translates into low thermal agitation and a collapse of the spin order. 
However, 2D systems with uniaxial magnetic anisotropy are able to withstand thermal agitation, allowing magnetic states in mono and multilayer materials.

In the past decades, the study of 2D magnetic properties has been  performed on epitaxially grown thin films, in which phenomena such as oscillating exchange coupling\cite{Nikolaev,Gareev}, giant magnetoresistance\cite{Baibich,Binasch} and Hall effect\cite{Li,Taylor} have been observed. 
Nevertheless, the study of the intrinsic magnetic properties of these 2D systems is novel, and most of the materials that have been synthesized are magnetic van der Waals crystals\cite{gong2017discovery,huang2017layer}.
In the last five years, non-van der Waals two-dimensional materials have also been synthesized, mainly by exfoliating naturally occurring ores.  
By liquid exfoliation of natural iron ore hematite ($\alpha$-Fe$_2$O$_3$), Balan et al.\cite{Balan-hem} synthesized a new 2D material which called hematene. Contrary to its antiferromagnetic bulk, hematene presents ferromagnetic order.  
Similarly, other promising material for 2D magnets is ilmenene \cite{Balan}, which has been synthesized using liquid phase exfoliation from  titanate ore ilmenite (FeTiO$_3$). 
Motivated by the synthesis of iron ilmenene, other similar 2D materials could be exfoliated from the bulk systems with their 3D counterpart ilmenite-like structures.

The aim of the present work is then to characterize the
ilmenene-like titanates (TMTiO$_3$, TM = V to Zn) to set a first theoretical basis for this exciting family of
materials.
Within the framework of the density functional theory, 
we systematically analyze the crystalline structure of these compounds, finding that most of the  ilmenenes
exhibit triangular symmetry for TMs on both sides of a Ti-Ti hexagonal graphene-like sublattice. 
In the chromium and copper ilmenenes, we find structural distortions of Jahn-Teller origin.
Our electronic structure calculations reveal that most of these compounds are magnetic semiconductors which have
TM layers antiferromagnetically coupled. The calculations including the spin-orbit interaction show a strong magnetic anisotropy, with the magnetization
being oriented out-of-plane (in-plane) below (above) half-filling of the TM electronic 3d shell. The presence of
out-of-plane anisotropy in some of these compounds suggests potential applications for spintronics in thin layers
and 2D materials.

%%%%%%%%%%%%%%%%%%%%%%%%%%%%%%%%%%%%%%%%%%%%% 
%%%%%%%%%%%%%%%% MODEL DETAILS %%%%%%%%%%%%%% 
%%%%%%%%%%%%%%%%%%%%%%%%%%%%%%%%%%%%%%%%%%%%%

\section{\color{red2}Model Details}

In this work, we 
study the structural, electronic and magnetic properties of ilmenene-type 
materials using the projector augmented wave method (PAW) implemented in the Vienna Ab-initio Software Package (VASP)\cite{VASP1,VASP2}. 
For the exchange and correlation potential we use the Perdew-Burke-Ernzerhof form of the generalized gradient approximation (GGA),  
with the formulation of Dudarev\cite{Dudarev} for GGA+U.  
The Hubbard U parameters for each element are chosen based on the available literature on transition metal oxides\cite{Aguilera-Ayuela_U}, 
and are shown in Table S1 of the Supplementary Information (SI) with the considered valence states. 
A test calculation of the partial density of states (PDOS) of cobalt titanate using the HSE06 hybrid functional 
shows that including a U parameter in the oxygen p-orbitals is needed to correctly describe the electronic structure. 
This finding is in agreement with previous investigations concerning titanium oxide and hematite\cite{Lany_2015,Lany, Morgan-PRB, Huang-Hematite, Jiang-PRL, Gou-PRB}.  
All calculations are performed with 
a well-converged plane-wave cutoff energy of 800 eV, a gamma-centered 4x4x1 Monkhorst-Pack k-point mesh, and a Fermi smearing of 20 meV. Atomic coordinates are relaxed until forces in all directions were smaller than 0.5 meV/\r{A}. An energy convergence criterion of 10$^{-7}$ eV is used. 
Further tests using even a larger plane wave cutoff up to 1000 eV and 6x6x1 k-point mesh do not modify the presented results. 
Differences in local charges and local magnetic moments are univocally analyzed using the Bader method\cite{Bader1,Bader2}. 
By including the spin-orbit coupling, additional tests are performed to converge the magnetocrystalline anisotropy energy with respect to the Brillouin Zone sampling.

%{\it Structural model.}
The d-metal ilmenene sheets are obtained by cutting their respective bulk titanates (TMTiO$_3$, TM = V to Zn) in the hexagonal [001] direction. 
For the iron ilmenene FeTiO$_3$, transmission electron microscopy measurements confirm the 2D structure in this direction\cite{Balan}.  
The layer structure is shown in Fig. \ref{figure1}.
For all the compounds under analysis, two different layer-ilmenenes are tested: 
titanium and transition metal terminated ilmenene layers. 
This work focuses on the TM ended layers because they are found to be more stable for all the materials under analysis.

\section{\color{red2}Results and discussion}

%%%%%%%%%%%%%%%%%%%%%%%%%%%%%%%%%%%%%%%%%%%%% 
%%%%%%%%%%% STRUCTURAL PROPERTIES  %%%%%%%%%% 
%%%%%%%%%%%%%%%%%%%%%%%%%%%%%%%%%%%%%%%%%%%%%

\subsection{Structural properties}

%\textit{Structural Properties.}

\begin{figure}[htb]
  \includegraphics[scale=0.16]{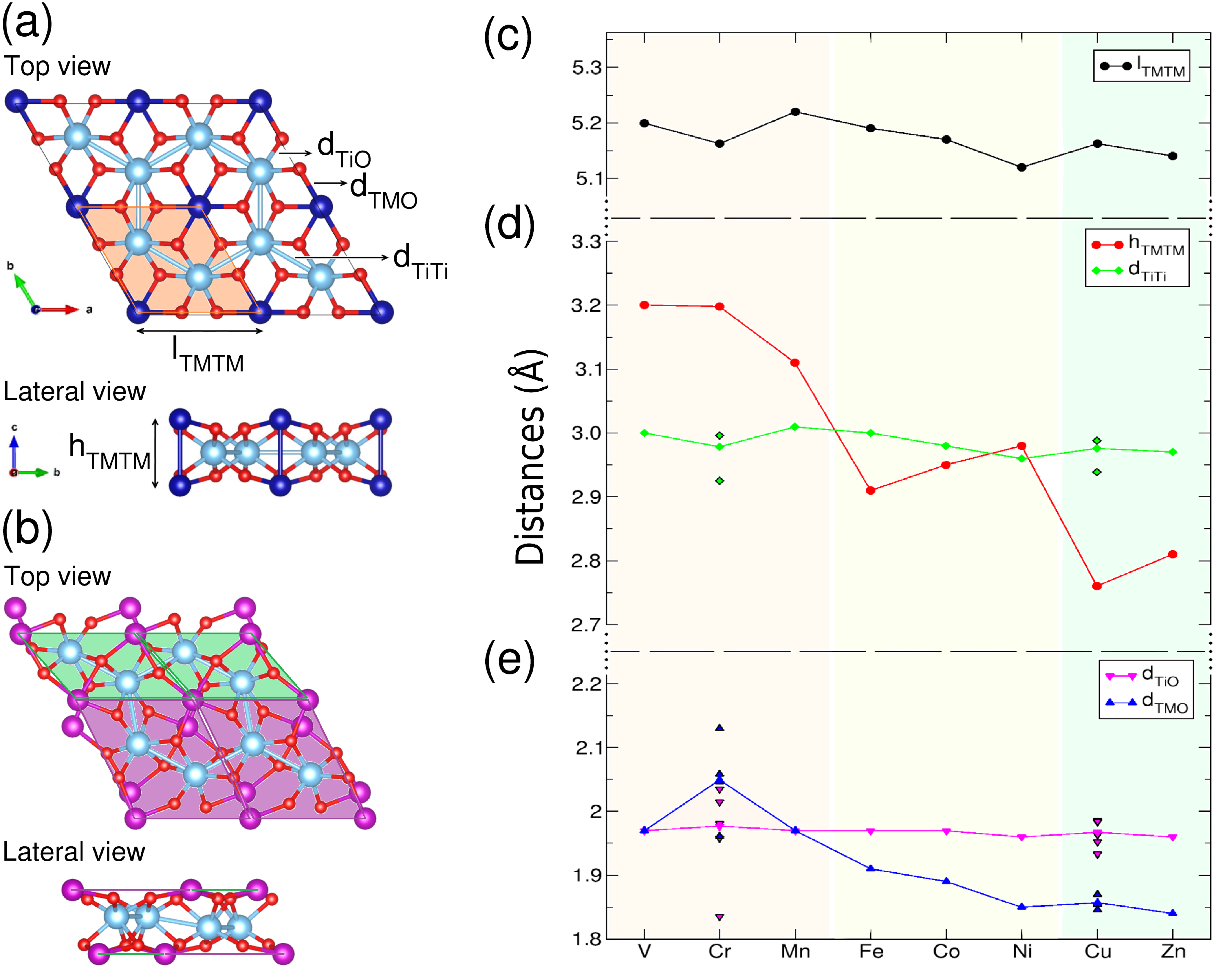}
  \caption{ %Example of the stable 
 Magnetic unit cell for transition metal ended 
  ilmenene-like systems: (a) symmetric for most ilmenenes, and 
  (b) distorted for chromium titanate CrTiO$_3$. 
  The color code of the atoms is as follows:   TM (Cr) atoms in blue (purple), titanium in cyan, and oxygen in red. The orange area in panel (a) represents the smaller chemical cell. 
  For chromium and copper titanates, the chemical and magnetic cells coincide.
  Calculated interatomic distances:  
  (c) horizontal TM-TM distance $l_{TM-TM}$, 
  (d) layer height $h_{TM-TM}$ and Ti-Ti distances, and 
  (e) Ti-O and TM-O distances. 
  For the chromium and copper ilmenenes we present an average of the distances in their distorted structures.
  }
  \label{figure1}
\end{figure}

The structure of the TM ilmenenes is graphically depicted from two viewpoints in Fig. \ref{figure1}. 
After structural relaxations, we find that most of the compounds keep the input symmetry, except for chromium and copper ilmenenes, which show  structural deformations of the perfect lattice due to the Jahn-Teller effect (see Fig. \ref{figure1}(b)). 
The orange area in panel (a) denotes the chemical cell that reproduces the crystalline structure of the ilmenenes when periodically repeated.  We calculate magnetic configurations using a larger 2x2x1 magnetic cell. 
For chromium and copper titanates, due to structural distortions, the unit cell and the magnetic cell coincide. 
In the distorted compounds, the two in-plane lattice vectors become slightly different: for the chromium case a=10.57\r{A} and b=10.00 \r{A}, and for the copper case a=10.44 \r{A} and b=10.21 \r{A}. Note that these ilmenenes become anisotropic. Overimposed on the global lattice distortions, the inner atomic distortions become more noticeable. 
For instance, the largest-smallest distances are 6.0 \r{A}- 3.9 \r{A} and 5.4 \r{A}- 4.8 \r{A} for Cr and Cu ilmenenes, respectively. 
The large 1D anisotropy of these two Janh-Teller distorted ilmenenes is shown by the stripes of the green and violet areas in Fig. \ref{figure1} (b).
A summary of the interatomic distances of the TM titanates is shown as a function of the elements across the period in the periodic
table in Fig. \ref{figure1}. 
We find that the horizontal distance between transition metals in the same layer $l_{TM-TM}$ increases
from V to Mn, then decreases until Ni, and finally increases for the brass metals Cu and Zn. 
The height distance between metals in different layers $h_{TM-TM}$ generally decreases. 
The distances involving Ti in the Ti-Ti and  Ti-O bonds remain nearly
constant, while the TM-O distances decrease. 
In essence, TM-ilmenenes can be thought of as a solid network 
formed by Ti and O atoms on which the TM atoms attach on both sides. 
For the Cr and Cu ilmenenes showing structural distortions, the TM-TM distances have two values because the triangular symmetry is broken, and the Ti-O distances are split into several values as shown by the extra marks in the lower panel of Fig. \ref{figure1}.

We find that in comparison with their bulk counterparts, 
TM ilmenenes are compacted along the c direction, so that a two-dimensional hexagonal layer of titanium ions is formed similar to graphene.
The Ti-Ti sublattice in ilmenenes
remains almost constant with the distances varying within 3\%.
\footnote{However, the chromium titanate layer does not show
the fully compacting behavior to a flat titanium sheet 
(see Fig. \ref{figure1}(b)).}
This compression for the iron ilmenene has a minimum height about 2.91 \r{A}, in agreement with another work \cite{Balan}.
The theoretical value of 2.59 \r{A}  for the interplanar space corresponding to the ($11\bar{2}0$) and
($\bar{2}110$) lattice spacing compares well with the experimental one ($\sim$ 2.53 \r{A}) \cite{Balan}, which is also supporting our calculations since the interatomic distances depend on the chosen U values.
For cobalt titanates, a layer thickness of 4.03 \r{A} has been reported\cite{CTO-Arruabarrena} in bulk, which decreases to 2.95 \r{A} in the ilmenene layer, while the horizontal distance between cobalt atoms in the same layer expands slightly from 5.06 \r{A} in bulk to 5.17  \r{A} in the layer. 
The significant changes in the height distances are due to the layer being decorated with half of the TM atoms than in the corresponding bulk, in order to keep the stoichiometry.   It seems that the ilmenene layers depart largely from those found in ilmenites and not only their structural properties but their magnetic ones are thus deserving a thoroughly study.

%%%%%%%%%%%%%%%%%%%%%%%%%%%%%%%%%%%%%%%%%%%%% 
%%%%%%%%%%% ELECTRONIC PROPERTIES  %%%%%%%%%% 
%%%%%%%%%%%%%%%%%%%%%%%%%%%%%%%%%%%%%%%%%%%%%
\subsection{Electronic properties: magnetic semiconductors}

\begin{figure}[htp]
    \centering
    \includegraphics[scale=0.42]{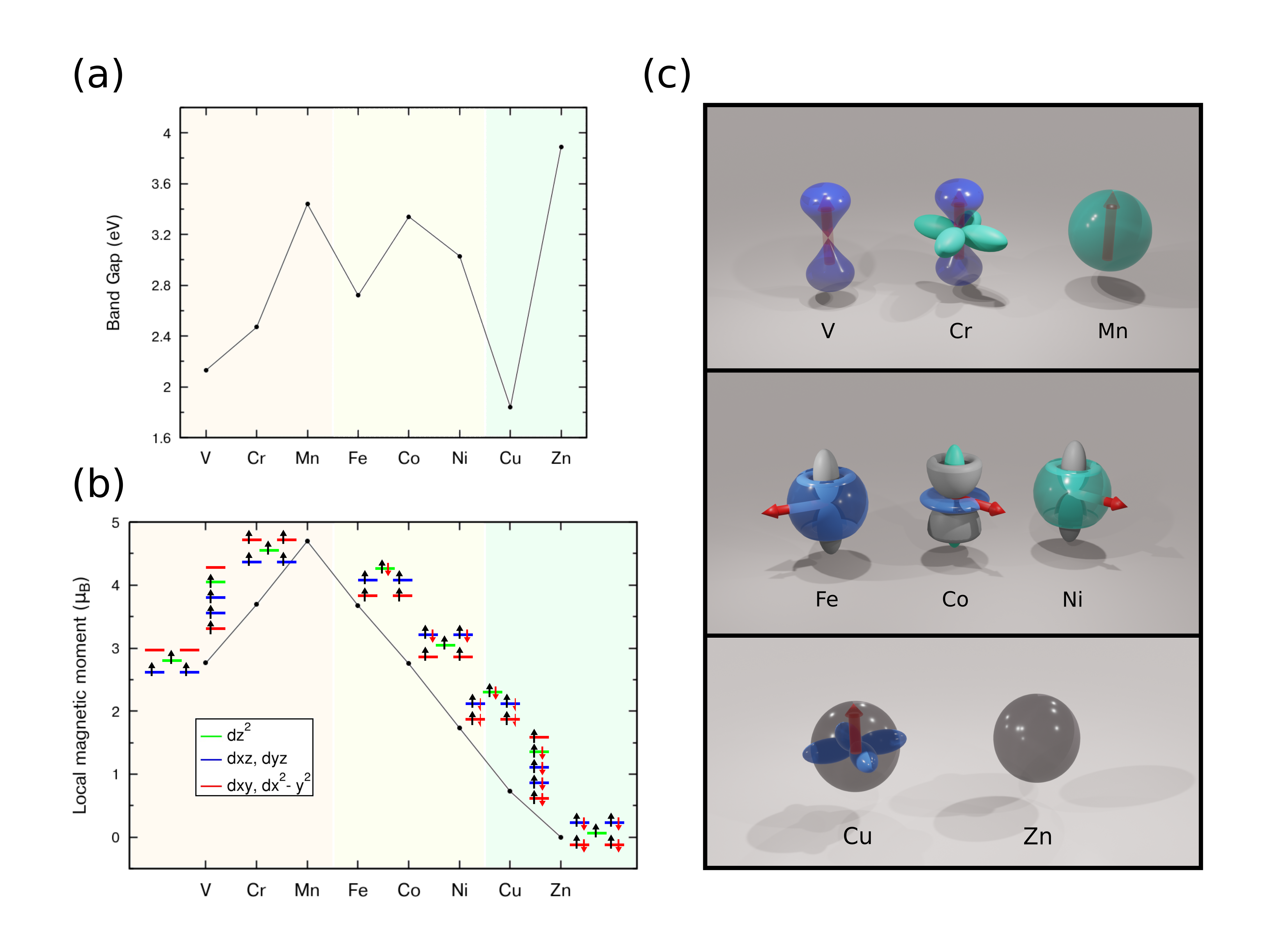}
    \caption{ (a) Electronic band gaps for TM ilmenenes. Vertical lines separate different regions with TM below half-filling, TM above half-filling, and brass metals. (b) Calculated local magnetic moment around transition metal atoms.  
    For each compound, the electronic filling model of the ground state is also shown. Red levels represent the in-plane $d_{x^2-y^2}$ and $d_{xy}$ orbitals; green, the out-of-plane $d_{z^2}$ orbital; and blue, the $d_{xz}$ and $d_{yz}$ orbitals with in- and out-plane components.
    (c) Orbital models of the local magnetization in 3d TMs and brass metals within ilmenenes. Blue and green regions denote spin polarized regions; grey, spin compensated ones. }
    \label{figure2}
\end{figure}

%\textit{Electronic Properties: Magnetic Semiconductors.}
We now focus on the electronic structure of ilmenenes and find that they show a gap opening that is linked to their stability.
The calculated electronic band gaps range between 1.8 and 4 eV, as displayed in Fig. \ref{figure2} (a), with values that are typical for semiconductors.
The gaps are increasing when going from V to Mn, and they are oscillating for Fe, Co and Ni. 
In general, the values are smaller than the gap for TiO$_2$ in the rutile bulk phase ($\sim$ 3.2 eV). 
Note that the Mn and the Co ilmenenes remain within the same order of the TiO$_2$ gap values.
These trends follow the filling of 3d TM electronic levels as discussed below.

The TM atoms in the ilmenene compounds show local magnetic moments, which are calculated using the Bader method,
and are shown in panel (b) of Fig. \ref{figure2}. The calculated values are close to the ones expected by
applying the Hund rules to isolated ions. Furthermore, the local magnetic moment as a function of atomic number
follows a Slater-Pauling-like rule, increasing from vanadium to manganese, then decreasing until it vanishes for
zinc. Thus, the 3d ilmenenes can be taken as magnetic semiconductors. 

%ELECTRONIC STRUCTURE 
The band structures and the PDOS projected onto the transition metal atoms are collected in 
the SI. 
From their analysis, an electronic filling model for each TM ilmenene is presented in Fig. \ref{figure2}(b). 
We find that the orbitals can be classified into three groups: (i) out-of-plane $d_{z^2}$, (ii) in-plane
$d_{x^2-y^2}$ and $d_{xy}$, and (iii) mixing in- and out- of plane contributions for $d_{xz}$ and $d_{yz}$. 
The $d_{x^2-y^2}$ and $d_{xy}$ orbitals are highly hybridized, as the $d_{xz}$ and $d_{yz}$ orbitals are. 
Because the in-plane degenerated energy levels of chromium and copper titanates are being partially occupied,
they are held responsible for the Jahn-Teller-like distortions in these structures, where a splitting of these
otherwise degenerated orbitals is observed under distortions. 

All these electronic trends can be better understood following the orbital model in real space for each case, as
shown in Fig. \ref{figure2} (c). Below half-filling, we show how the magnetic moment increases  in correlation
with the gap values. For next elements, above half-filling, the Fe case becomes similar to the Ni one which has a
weaker spin distribution. 
In between, we find the Co ilmenene with a large in-plane spin distribution has also an out-of-plane component that cannot be neglected, an extra component that is related to the complexity added to this case when trying to write down a single spin hamiltonian. 
For brass metals, the local magnetization is nearly screened having a spin compensated cloud around the brass metal centers. This model is going to be interesting in the discussion of magnetic anisotropies below. 
For instance, we note that the spin distribution is not anisotropic for the case of Cr (Cu) having one-electron less below a half-filled (filled) 3d shell.

%%%%%%%%%%%%%%%%%%%%%%%%%%%%%%%%%%%%%%%%%%%%% 
%%%%%%%%%%%%% MAGNETIC ORDERING  %%%%%%%%%%%% 
%%%%%%%%%%%%%%%%%%%%%%%%%%%%%%%%%%%%%%%%%%%%%
\subsection{Magnetic ordering}

\begin{figure}[htb]
  \includegraphics[scale=0.23]{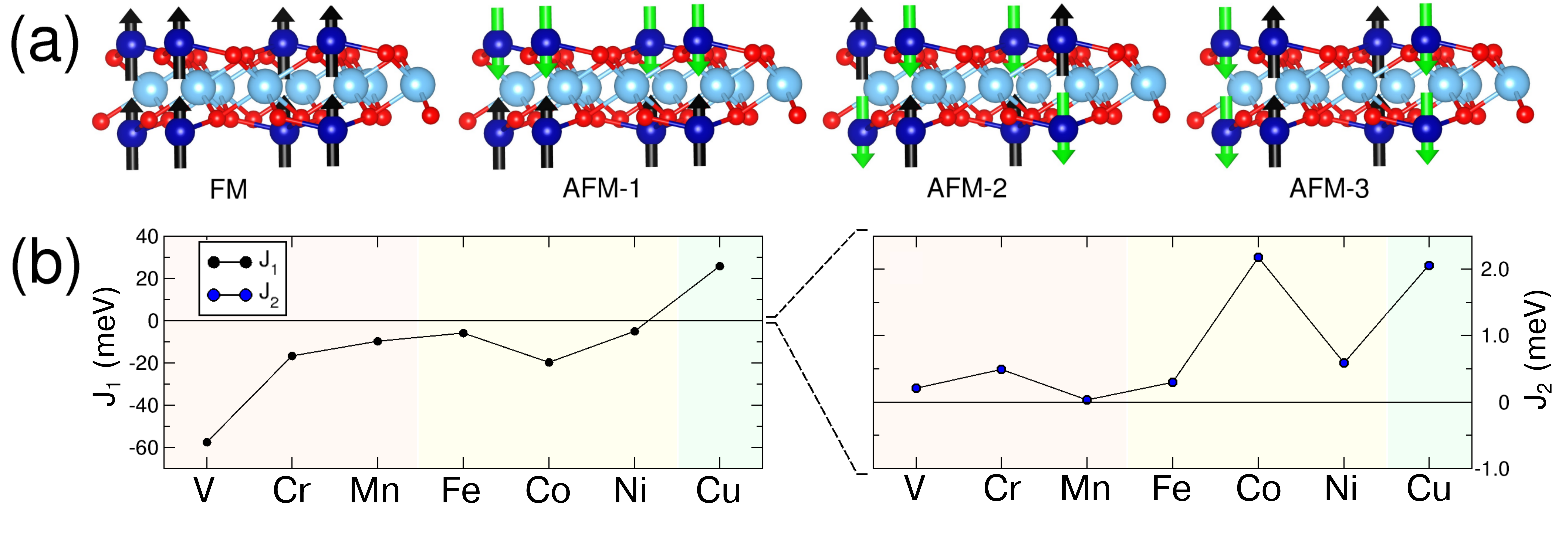}
  \caption{ (a) Magnetic ordering configurations of TM ilmenenes: ``FM'' Ferromagnetic, ``AFM-1'' antiferromagnetic by layers, ``AFM-2'' and ``AFM-3'' antiferromagnetic.
  (b) Couplings between both layer sides (J$_1$), and within the same side layer of TM ions (J$_2$).}
  \label{figure3}
\end{figure}

%\textit{Magnetic Ordering.}
We next consider the magnetic behavior of the ilmenene-type materials, and calculate the total energies of the magnetic configurations  shown schematically
in Fig. \ref{figure3}.

We find that, with the exception of ferromagnetic copper and spin-compensated zinc titanates, the ilmenenes show
antiferromagnetism between the 2D TM layers in the so-called AFM-1 configuration.

The magnetic configurations in Fig. \ref{figure3} are used to fit the Heisenberg Hamiltonian 
\begin{equation}
H = -\sum_{ij}^{} J_{ij} \mathbf{\tilde{S}_{i}} \cdot \mathbf{\tilde{S}_{j}} 
\end{equation}
where the $\tilde{S}_i$ is the pseudospin of each isolated atomic species (e.g: $\tilde{S}=3/2$ for Co), and the $J_{i}$ with $i= 1, 2$ are the inter-layer ($i=1$) and intra-layer ($i=2$) magnetic couplings schematically depicted in \ref{figure3}. 
The magnetic couplings are computed from the energy differences $\Delta E_1 =  E_{FM} - E_{AFM_1}$, $\Delta E_2 =  E_{AFM_2} - E_{AFM_1}$. 

The fitted J$_{i}$ values are shown in Fig. \ref{figure3} (b). 
The inter-layer coupling is the leading interaction because J$_1$ is an order of magnitude larger than J$_2$. 
The inter-layer coupling J$_1$ is negative for most of the ilmenenes, that indicates a preference for antiferromagnetism. 
The intra-layer coupling J$_2$ is positive in all cases, and the compounds favor intralayer ferromagnetism. 
Since the coordination of the TM atoms on each layer side changes from being hexagonal in bulk to triangular in the layer, the structural differences between bulk and two dimensional titanates mentioned above play an important role in the magnetic ordering. 
It is noteworthy that the manganese ilmenene even favors ferromagnetism within layers in contrast with its antiferromagnetic bulk counterpart.

%%%%%%%%%%%%%%%%%%%%%%%%%%%%%%%%%%%%%%%%%%%%% 
%%%%%%%%%%%% MAGNETIC ANISOTROPY  %%%%%%%%%%% 
%%%%%%%%%%%%%%%%%%%%%%%%%%%%%%%%%%%%%%%%%%%%%
\subsection{Magnetic anisotropy}

\begin{figure}[htp]
  \includegraphics[scale=0.3]{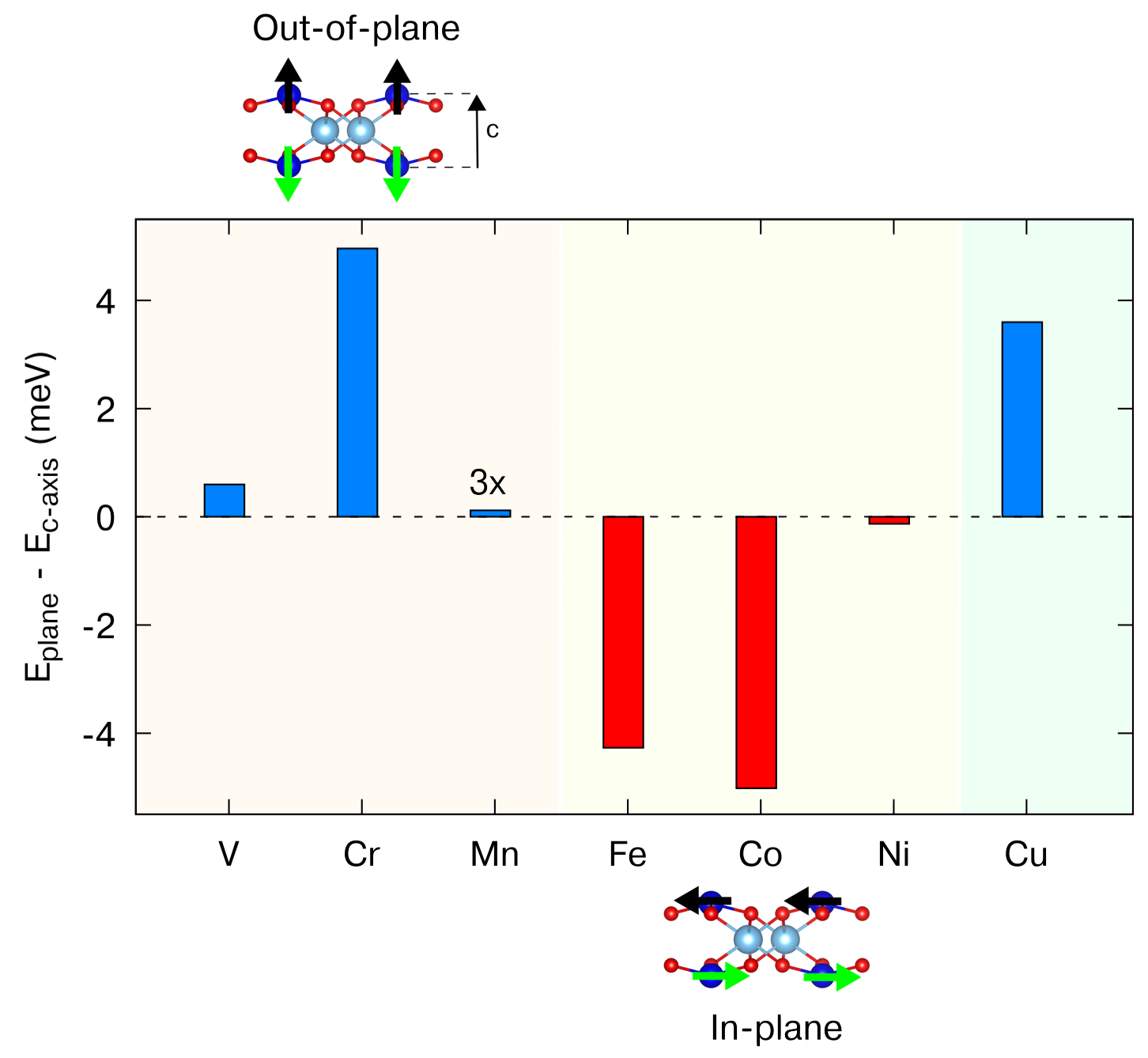}
  \caption{Magnetocrystalline anisotropy energy of the transition metal ilmenenes. In blue, compounds with out-of-plane anisotropy; in red, with in-plane magnetic moments. For the case of manganese titanate, the magnitude was enhanced by a factor of 3 to make it visible in the shown range.}
  \label{figure4}
\end{figure}

%\textit{Magnetic Anisotropy.}
We last study the magnetic anisotropy of d-metal ilmenenes. 
The spin-orbit interaction couples the spin magnetic moment to the crystal lattice, which means that some spin
orientations are more stable than others. 
The magnetocrystalline anisotropy energy (MAE) 
is defined as the energetic difference between two magnetic configurations with different spin orientations. 
We calculate the total energy of the ilmenenes for a number of spin orientations, 
and find that, anisotropy-wise, ilmenenes can be classified in two groups: 
out-of-plane ilmenenes, in which the spin magnetic moment is aligned in the \textit{c}-axis, and in-plane ilmenenes, with the spin aligned 
in any of the directions along TM-Ti bonds projected in the plane  
(see Fig. \ref{figure4}). 
Due to symmetry around TM atoms, there are six such equivalent directions in the \textit{ab} plane.  
The vanadium, chromium and manganese titanates in addition to copper ones have an out-of-plane anisotropy, and the magnetization in iron,
cobalt and nickel titanates is oriented in-plane. These MAE trends agree with the electronic filling model, in which
below half-filling TMs the electrons near the Fermi level are those that include an out-of-plane component, and for iron, cobalt and
nickel titanates they are mainly in-plane.

The obtained MAE values range between several orders of magnitude from 10$^{-2}$ meV up to tenths of meVs because the spin-orbit interaction varies greatly for the different ilmenenes. 
The chromium, iron and cobalt titanates have a strong MAE around 5 meV, and in the case of the vanadium and nickel layers, this value is still large about 0.6 and 0.13 meV, respectively. 
The manganese titanate has the smallest anisotropy ($\sim$ 0.04 meV) because the orbital magnetic moment is nearly fully quenched.
This finding summed to having the energetic difference between the magnetic AFM-1 and AFM-2 configurations of the manganese titanate being very small (around 0.25 meV) points to non-collinear magnetic configurations, 
a topic that might merit future experimental investigation on this specific layer, but being beyond the actual scope of the paper. Furthermore, the iron ilmenene is also a particular 2D layer because it shows a strong in-plane anisotropy, contrary to the ilmenite bulk with a non-negligible out-of-plane component.

We observe a noticeable trend for the magnetic anisotropy. Below half-filling, the V and Cr ilmenenes exhibit
out-of-plane anisotropy, while the half-filled manganese ilmenene has a very small MAE. Above half-filling, most
the compounds have an in-plane anisotropy. In fact, these trends in anisotropy follow the levels depicted in Fig.
\ref{figure2} (c). For V and Cr ilmenenes, it shows that the spin density is  perpendicular to the plane. 
The spin density in manganese ilmenene is isotropic, and this correlates with its low MAE. 
Above half-filling, the spin density of Fe, Co and Ni ilmenenes lies in-plane, in agreement with their in-plane
MAE. The case of Cu ilmenene shows out-of-plane anisotropy even if the spin density lies in-plane, a finding that
can be explained by the out-of-plane changes in the orbitals when the spin-orbit coupling term is included with
the distortions.

We next bring the magnetic results on magnetic ilmenenes into contact with experimental facts. Note that Fe
ilmenenes  can be obtained by exfoliation and deposited on substrates for some experimental setups \cite{Balan, Balan-hem}. Then, further measurements on the magnetic properties of the Fe ilmenenes can be today performed in
line with our theoretical results.
Furthermore, we have studied Co ilmenenes that could be interesting as they can behave in a similar way to study
magnon physics in two dimensions as their bulk counterpart is already being used
\cite{CTO_Nat.Comm.,CTO-PRX,CTO-Arruabarrena}. 
In fact, Cr ilmenenes seem to become key as they are candidates to show out-of-plane anisotropy when in the form
of ultrathin layers. These findings are suggesting the future use of TM ilmenenes in spintronics devices for
injecting magnons and study magnetism in exfoliated 2D magnetic layers.

\section{\color{red2}Conclusion}
In this work, we have analyzed the structural, electronic and magnetic properties of TM ilmenene-like systems.
Our calculations reveal that most of the materials under analysis present a triangular crystalline structure for TMs, with an ironed compression of the internal titanium ion layer with respect to bulk.
The chromium and copper 
ilmenenes exhibit notable structural distortions, which seem to have a Jahn-Teller origin. 
All the compounds studied have been found to be magnetic semiconductors with band gaps in the range between 1.7 and 4 eV.
The magnetic ground state is mainly antiferromagnetic between layers, and ferromagnetic and spin-compensated for CuTiO$_3$ and ZnTiO$_3$, respectively. 
Furthermore, spin-orbit calculations revealed that TM ilmenenes can be divided into two  groups: 
out-of-plane ilmenenes, for less than half-filling of the 3d bands, 
and in-plane ilmenenes, above half-filling of these levels, with the spin aligned in the TM-titanium directions projected in the hexagonal plane. 
We believe that this family of materials paves the way to other types of promising 2D candidates with potential applications in the field of spintronics. 

\section{Data availability statement}

All data that support the findings of this study are
included within the article (and any supplementary
files).

\section{Acknowledgement}

This work has been supported by the Spanish Ministry of
Science and Innovation with PID2019-105488GB-I00 and
PCI2019-103657. We acknowledge financial support by the
European Commission from the NRG-STORAGE project (GA 870114).
The Basque Government supported this work through Project No.
IT-1569-22. M.A was supported by the Spanish Ministry of
Science and Innovation through the FPI PhD Fellowship
BES-2017-079677. R.H.A-T acknowledge the postdoctoral contract
from the Donostia International Physics center.

\section{\color{red2}References}

%\bibitem{ilmenene.bib}
\bibliographystyle{iopart-num}
\bibliography{ilmenene.bib}

\end{document}